# MATHEMATICS AND INCENTIVES IN THE SLUMS

G. Stefansson[1], J. Lentin[2], A. H. Jonsdottir[3], E. Eiríksson[3], A. Kristjánsson[3]

[1]University of Iceland, Science Institute (ICELAND)
[2]Shuttle Thread Ltd (UNITED KINGDOM)
[3]Háskóli Íslands (ICELAND)

## Abstract

In response to COVID-19, a new project was started to allow students to solve computerised math drills outside of school. In 9 months it has gone from zero to one thousand students, in co-operation with ten community libraries in various slums and low-income regions in Kenya. The program uses the tutor-web as a study environment and access is provided by donating tablet computers to participating community libraries. Students are rewarded using the SmileyCoin cryptocurrency as they progress through the system and the libraries are free to sell for SMLY small food items, sanitary pads and the tablets themselves. The reward system is designed to put an emphasis on secondary school mathematics. Completion of the corresponding collection of drills gives SmileyCoin awards sufficient to purchase a tablet.

Conclusions based on the first year indicate that the resulting effect on participation and performance is unprecedented: Eleven libraries with 1301 students opted for voluntary participation in 2021 causing the program to run at full financial capacity. In that year, 450 students earned enough SMLY to purchase the tablets, which involves completing a large collection of drills to a level of excellence.

Status exam questions are dispersed within the drill collection. These independent measurements show learning which surpasses both rote learning and mechanical learning and demonstrate an increase in the general ability to address new mathematical problems not seen before.

## 1   INTRODUCTION

The tutor-web[1] system is designed for research[2] and learning[3]. Its drills are primarily used for learning so there are typically no limits on the number of attempts at improving performance. An important feature is that for most drills the student is shown a detailed explanation of the solution immediately after choosing an answer option. This system is used at multiple schools and universities in Iceland and Kenya, mostly for mathematics and statistics. Students earn SmileyCoin, a cryptocurrency, while studying [4].

In Kenya the system is a part of a plan to enhance mathematics education using educational technology, organised by the Smiley Charity in collaboration with the African Maths Initiative and several individuals. This had been implemented by donating servers to schools and tablets to students. Typically these schools did not have Internet access so the cryptocurrency could not be used.

Schools were closed in Kenya in 2020 so agreements were set up with community libraries instead.

This paper describes how this unintended experiment has affected the development and use of the tutor-web.

### 1.1   Redesigning the tutor-web project due to COVID-19

The single most important learning goal of this project in Kenya is to assist secondary school students with getting into university. The most difficult hurdle in this regard is passing the national exam, known as the KCSE. The tutor-web has a generic tutorial (module) for secondary school mathematics but in 2020 a new module was added, specifically targeting the KCSE. This was implemented by taking questions from an example exam and writing a computer program for each question. The program was run to obtain drill sets of 100 drill items for each question and the entire collection of drill sets was installed as a module in the tutor-web. This takes into account problems with rote learning[5].

A non-profit organisation, the Smiley Charity (registered in Iceland as Styrktarfélagið Broskallar), already ran a project to donate servers to schools and tablets to students, designing servers to run without any Internet connection and hand-delivering tablets to students to emphasise their personal ownership of the tablets. The project ran under the name Education in a Suitcase since normally the tablets and servers have been brought as hand luggage to each site.

As described in more detail elsewhere[6], considerable redesign was needed during COVID-19 since schools in Kenya remained largely closed so tablets could not be donated directly to students. Community libraries in Kenya remained open during the COVID-19 pandemic and became a place for students to come in to study. The Smiley Charity therefore arranged with several partners in Kenya to distribute tablets to select libraries. This is a considerable change in strategy since the tablets no longer go directly to students from the charity. Instead, they are purchased by partners in Kenya and donated under certain conditions to partnering libraries.

Naturally there are pros and cons to each method of giving tablets directly to students or to an organisation. In the former case the recipient and end user are well defined. However in the latter case the tablet may find much more use since it is borrowed by more students.

The tutor-web has rewards in terms of grades, like other drilling systems, but it can also reward students with an electronic token, the SmileyCoin. The SMLY is a cryptocurrency, which can be redeemed by the student and used outside the tutor-web. The SmileyCoin Fund is an organisation which holds a large number of SmileyCoin. The fund accepts SmileyCoin grant applications and has funded the tutor-web with 1-1.5 bn SmileyCoin annually in recent years.

The schools and students, which the Smiley Charity has worked with up through 2020, have not had wireless connections and usually not even Internet. In these cases the blockchain is not accessible and SmileyCoin reward schemes have not been an option.

The cooperating libraries all have local WiFi and access to the Internet. The students are therefore able to earn SmileyCoin rewards as they progress towards high grades in the tutor-web. The rewards are not dependent on prior knowledge but on studying and solving exercises persistently enough to eventually obtain a high grade, which only depends on recent answers.

Each library can start lending tablets to students based on much fewer donated tablets than corresponding to a regular classroom of 20-30 students. Initially therefore, each library receives only 5-15 tablets.

The primary purpose of the SmileyCoin is to encourage learning. Further, for a reward scheme to have an impact, the reward has to make the effort required worthwhile. In earlier SmileyCoin experiments, small on-line markets have been set up where the students can purchase items such as discount coupons or tickets to the cinema. In the library model a different option opens up, which is to allow the diligent student to purchase the tablet itself using SmileyCoin earned by studying.

The tutor-web has two student groups, self-registered students and students who are registered as a part of a real-world classroom (or library). Reward schemes are set differently for the two groups, to avoid abuse of the cryptocurrency.

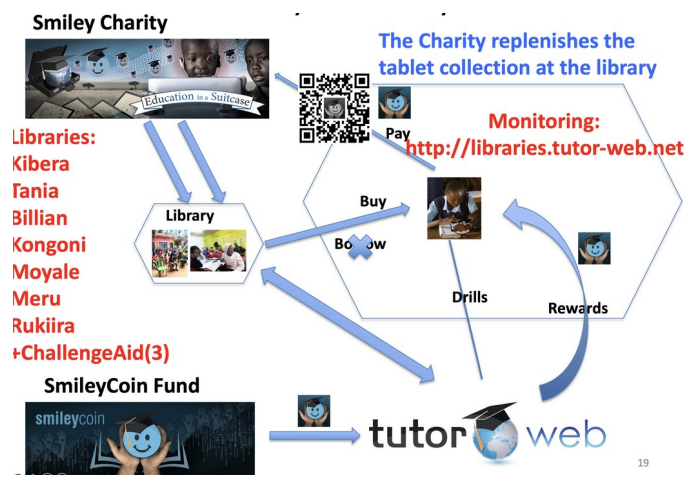

*Figure 1. The library model, as implemented in 2021.*

These considerations lead to the **library model**: The tablets are donated to a library, which has full discretion of how the tablets are lent to students. Pre-registered student accounts are set up in the tutor-web and assigned to students by the librarian. Each student can then practise for as long as they like, including towards acing the complete KCSE drill set. Acing the KCSE drill set gives a predefined reward of 1 M SMLY. On the back of each tablet is a QR-code corresponding to a payment address. If a student scans this code they get an option to pay for this tablet with 1 M SMLY. If a library gets to the stage of students purchasing their tablets, the Smiley Charity replenishes the tablet stock. Figure 1 gives a schematic of how the "library model" works.

Some care is needed in tuning the rewards in this scenario. The SmileyCoin Fund has finite SmileyCoin resources and the Smiley Charity has finite fiat funds and the rewards can be tuned at several scales so there are a fairly large number of options on how to assign rewards for performance.

A number of issues had to be solved before the move to libraries could be undertaken, including a change in the contract with the funding agency (Ministry of Foreign Affairs in Iceland), setting up agreements with libraries and arrangements with contact persons and NGOs in Kenya. Videos explaining the new approach were generated (e.g. https://bit.ly/TheLibraryModel) and the reward settings in the tutor-web were modified.

Data on the answers to drill items are collected in a tutor-web database. Normally these are used by instructors to monitor their students but in the present setting there are no instructors. These data are therefore only accessible for general anonymised statistical analyses.

Monitoring scripts were also set up, anonymous in terms of students, yet giving incentive for some competition amongst libraries. The output from the scripts is placed on a web page at https://libraries.tutor-web.net/.

## 2 RESULTS

### 2.1 The response

Figure 2 gives a summary of the response to the programme. Initially, while new libraries were being recruited, accounts were issued liberally to libraries, typically 100 accounts at at time. Adoption was initially very slow as seen in the figure but increased dramatically after that.

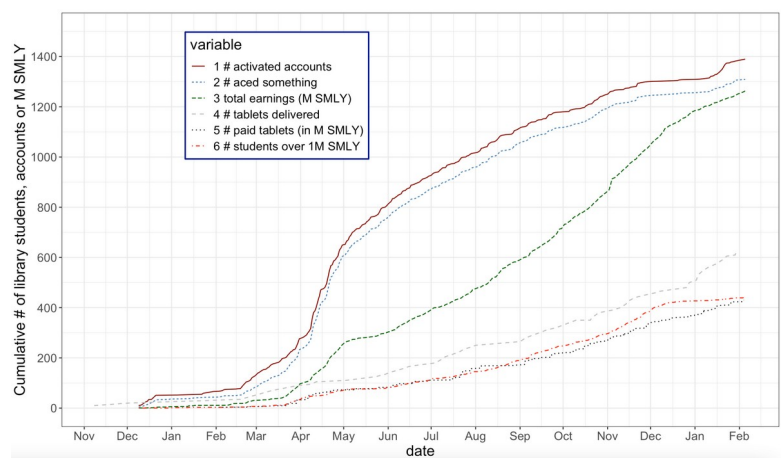

*Figure 2. Cumulative tutor-web use from 2020 into 2021, earnings and tablet purchases.*

Several time series are shown in the figure, the top curve (1) being the cumulative number of students who have accessed the tutor-web system and answered at least one question. Only slightly lower is the curve describing the cumulative number of students (2) who have completed at least one drillset to perfection. The green curve (3) shows the cumulative grand total number of SmileyCoin earned by library students.

Table 1. Activity by library.

|  | Students with 1M SMLY | Sold tablets | Delivered tablets |
|---|---|---|---|
| BillianSoM | 3 | 7 | 15 |
| KCRCLib | 8 | 1 | 15 |
| knlsKibera | 53 | 57 | 75 |
| knlsMeru | 48 | 38 | 60 |
| knlsMoyale | 149 | 177 | 195 |
| KongoniLib | 64 | 56 | 65 |
| MadoyaSoH | 4 | 7 | 25 |
| MajengoSoH | 4 | 7 | 20 |
| MashimoniSoH | 55 | 67 | 85 |
| MathareMCE | 0 | 0 | 15 |
| RukiiraLib | 3 | 4 | 20 |
| TaniaCentre | 50 | 9 | 15 |
| Total | 441 | 430 | 605 |

The bottom three curves relate to tablet accounting, indicating the number of tablets delivered to libraries, the number of tablets bought by students and the number of students who have earned at least the one million SMLY needed to purchase a tablet.

## 2.2 Budgets and controls

The programme has two distinct budget limitations, one in the number of SmileyCoin made available to the tutor-web system. This is donated through a grant by the SmileyCoin Fund and has been limited to some 1.5 bn SMLY per year. The other limitation is the financial budget of the Smiley Charity, which limits the number of tablets shipped to libraries to a total of 10-15 per week. Several controls can be used to ensure that the program stays within limits. The main controls are the number of tutor-web accounts issued and the number of tablets issued to libraries.

In addition to these controls, which relate directly to finances, the tutor-web system can be tuned in a variety of different ways. Students download drills to their devices and answer these in some order, items being replenished if the student remains on-line. Issues affecting system performance include the available number of items in a drillset, the number downloaded, how the difficulty of allocated items varies as a function of the student's grade and so forth.

The exact formulation of drills is also important. Initially all the secondary school drills were based on a single exam (a KCSE). It was subsequently found that some of these exam questions were far too easy or far too difficult. Therefore new batches of drills were added to these drillsets. Some of these were deliberately easy whereas others were made harder.

Table 2. *Frequency of aced KCSE drillsets by library: Number of students who have aced the indicated number of drillsets.*

|  | Number aced | | | | | |
|---|---|---|---|---|---|---|
|  | 1-20 | 21-35 | 36-50 | >50 | | Total |
| Library |  |  |  | earlier | This month |  |
| BillianSoM | 55 | 1 | 0 | 0 | 0 | 56 |
| KCRCLib | 11 | 4 | 1 | 4 | 0 | 20 |
| knlsKibera | 68 | 39 | 17 | 9 | 1 | 134 |
| knlsMeru | 16 | 1 | 2 | 38 | 7 | 64 |
| knlsMoyale | 151 | 57 | 31 | 87 | 6 | 332 |
| KongoniLib | 30 | 7 | 4 | 56 | 1 | 98 |
| MadoyaSoH | 16 | 2 | 0 | 0 | 0 | 18 |
| MajengoSoH | 14 | 0 | 0 | 1 | 0 | 15 |
| MashimoniSoH | 127 | 5 | 3 | 27 | 5 | 167 |
| MathareMCE | 25 | 0 | 0 | 0 | 0 | 25 |
| RukiiraLib | 8 | 0 | 0 | 2 | 0 | 10 |
| TaniaCentre | 74 | 37 | 11 | 38 | 0 | 160 |
| Total | 595 | 153 | 69 | 262 | 20 | 1099 |

At the time of the greatest increase in student participation the opportunity was used to deliberately include items extended the national curriculum. The logic behind this is that a tablet computer is a very expensive device and should only be awarded if the student has put considerable work into their studies. These additional drills were set up in such a manner as to be self-explanatory but requiring new thinking. If the student answered incorrectly, a detailed answer was provided, which should help the student in addressing the next item from the same batch.

## 2.3 The response

The activity in the individual libraries varies quite a bit, as seen in Table 1. It should be noted that some have been active for the entire period while others entered late in the year.

## 2.4 Progress

Progress can be measured in multiple ways, as grades, completion of drills, tablet purchases or performance on independent exams which are not a part of the drillsets proper.

There are over 50 drillsets for the secondary schools, each with from 100 to several hundred drills.

The tabular setup in table 3 can be used to show how far the student group in each library has progressed through a collection of secondary school drillsets. The table shows the number of students that have completed a given number of drillsets with excellence ("aced them", i.e. obtained a grade of at least 97.5%). For example, 68 students in Kibera are in their starting phase, having completed between 1 and 20 drillsets. In Meru 38 students had completed more than 50 drillsets and earned 1M SMLY before the start of the current month. Twenty students - in Kibera, Meru, Moyale, Kongoni and Mashimoni - were active in the given month, having recently completed the entire collection.

It is also of interest to see the time development of progress through the drillsets, as given in Table 3.

Table 3. Time development of aced KCSE drillsets: Number of students who had aced the indicated number of drillsets before the indicated date.

| Number aced<br>EndDate | 1-20 | 21-35 | 36-50 | >50 | Total |
|---|---|---|---|---|---|
| 2020-12-01 | 0 | 0 | 0 | 0 | 0 |
| 2021-01-01 | 11 | 0 | 0 | 0 | 11 |
| 2021-02-01 | 21 | 0 | 0 | 0 | 21 |
| 2021-03-01 | 29 | 1 | 0 | 0 | 30 |
| 2021-04-01 | 52 | 6 | 0 | 0 | 58 |
| 2021-05-01 | 193 | 90 | 33 | 27 | 343 |
| 2021-06-01 | 268 | 116 | 47 | 34 | 465 |
| 2021-07-01 | 327 | 119 | 50 | 51 | 547 |
| 2021-08-01 | 357 | 123 | 53 | 59 | 592 |
| 2021-09-01 | 385 | 127 | 55 | 64 | 631 |
| 2021-10-01 | 418 | 129 | 55 | 73 | 675 |
| 2021-11-01 | 450 | 132 | 57 | 88 | 727 |
| 2021-12-01 | 484 | 134 | 60 | 133 | 811 |
| 2022-01-01 | 529 | 139 | 63 | 178 | 909 |
| 2022-02-01 | 584 | 150 | 66 | 262 | 1062 |
| 2022-03-01 | 595 | 153 | 69 | 282 | 1099 |

## 2.5 Abuse

This paper describes work using incentives which may not be directly financial, but can be made so by using cryptocurrency exchanges where individuals can exchange SmileyCoin for Bitcoin or fiat currencies such as US dollars or Kenyan shillings. Several examples of abuse were detected during the first year of the library project. Most notably, several usernames were clearly used by individuals who were not students. These users answered all questions correctly every time and did so in an unreasonably short amount of time. In one instance the abusers turned out to be university students who had gained access to the usernames in one of the libraries. In other instances the abuse was clear from the data, but the perpetrators were not found. This has no major impact on the project as a whole, but needs to be kept in mind when interpreting some of the results. Note that the abusers may earn SmileyCoin but will not receive tablets unless a librarian is a part of the abuse.

## 2.6 Monitoring more than the simplest progress

The tutor-web is primarily a system for learning rather than evaluation. However, an important aspect of learning is self-monitoring and the system therefore produces a grade which is based on a weighted average of recent answers. In this manner, the student can monitor progress, the instructor can use the same measure as a partial course grade and the tutor-web has a measure which can be used for SmileyCoin rewards. Several of the references[1-6] describe different aspects of this grading scheme and demonstrate the improvement in the students' performance as they progress through the system.

The analyses in [5] demonstrate one of the difficulties in self-learning systems like the tutor-web, where the user can continue working the system until they have seen most of the questions and simply see them again. It is clear that many users have a tendency to work in this way and anonymous tips also indicate that this is a problem. This problem is alleviated in the library setup, however, where the drill items are generated in very large numbers and are generated using random numbers to avoid direct repetition. The earlier analyses[5] also show that even when there is considerable rote learning, the students' performance is also enhanced with practise, on items not seen before, so there is indeed also meaningful learning.

What remains is the issue of whether this learning only leads to (meaningful) learning on items of the same form, e..g if the student is practising differentiation, does that only lead to learning the methods of differentiation or does the act of practising mathematics also lead to a general increase in ability to work out unseen math problems?

To investigate this question, an 18-question status exam was used. Each of these questions was placed in every single secondary school drillset. The typical student is expected to start with the first drillset, which has some 544 drills. The student may quickly complete a dozen items correctly and proceed to the next drillset. Alternatively, the student may struggle and need to answer a hundred items before proceeding. However, with each requested drill item, the student will receive a status exam question with some positive probability. As students progress through the drillsets, they will see a status exam question from time to time. From the start of this experiment through January 2022, status exam question have been allocated almost 200 thousand times (out of some 3 million drill item allocations).

Considerable care is needed to interpret these data. For example, the student may see the same status exam question several times, in different drillsets. Some students may also give up before completing all drillsets. Incorporating all these effects into a generalised linear model demonstrates highly significant student effects as well as the importance of repetition and the difference in the difficulty of the status exam questions. But most importantly, there is a highly significant effect of where the student sees the question, i.e. there is a significant difference in performance depending on in which drillset the student gets the status exam item.

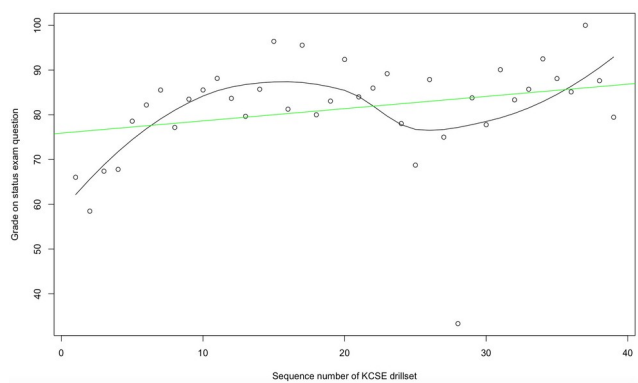

*Figure 3. Grade on first-time-seen status exam questions as a function of the ordered drillset in which the question was seen. Points are average grades across status exam questions received within the ordered drillsets. The line is a simple linear regression whereas the curve is a loess smoother.*

To minimise the effects of such biases and produce a simple yet descriptive plot, the data are reduced to analyse only the first time each student sees a status exam question. Further, dropouts deleted from the analysis and only those students are included who have both completed the entire secondary school curriculum to excellence and seen all 18 status exam questions.

These result, given in Figure 3, therefore addresses the following question: **Considering only those students who completed all drillsets, to the level of earning a tablet, how does their performance vary as they proceed through all the drillsets?** Notice the outlier in drillset 28, which pulls down the slope. The point estimate of the slope corresponds to an increase of 12 percentage points on the status exam.

Note the general variability in the point cloud: There is no particular reason for a large difference between adjacent points, except for natural variation due to low sample sizes. One would expect the points to move together as the sample size goes up.

Note also that those abusers mentioned above who know the entire material typically log in as multiple users and always answer correctly, including on every status exam question. They will therefore correspond to a straight line at 100% in Figure 3, biasing downwards the point estimate of learning.

It will be difficult, but not impossible, to distinguish between the abusers and the students who simply know the material. One indication is in whether these users sent their SmileyCoin to a cryptocurrency exchange instead of purchasing a tablet computer.

A different approach is to just focus on students who clearly did not know the material when they started. This will address a slightly different question from Fig 3, namely: **Does practise towards excellence in the tutor-web provide a struggling student with increased general ability in mathematics?** This question is even more relevant than the corresponding question for the generic student, as address in Fig. 3.

## 3 CONCLUSIONS

A method is described to enhance mathematics education in Kenyan slums by providing tablets to community libraries and allowing students to purchase the tablets by using a cryptocurrency earned by studying. The resulting effect on participation and performance is unprecedented: Eleven libraries with 1301 students opted for voluntary participation in 2021 causing the program to run at full financial capacity. In that year, 450 students earned enough SMLY to purchase the tablets, which involves completing a large collection of drills to a level of excellence.

While practising in the tutor-web drilling system, the students demonstrate an increased ability in general mathematics ability, measured by a yardstick which is not a part of the drills themselves.

These results are preliminary and will be sharpened through more detailed analyses and increased data collections in 2022.

## ACKNOWLEDGEMENTS

A large number of individuals and institutions have made this work possible. Through the years, the projects have received funding from The Icelandic Centre for Research and from several EU H2020 grants, most recently FarFish (Horizon 2020 Framework Programme Project: 727891 — FarFish). The current initiative in the Kenyan libraries is funded by the Icelandic Ministry of Foreign Affairs, including tablet purchases and corresponding logistics.

Continuous support has been provided by the University of Iceland where the course material has been developed, and by the University of Iceland Science Institute where most of the research and development has been conducted. The current version of the tutor-web was developed by Jamie Lentin at Shuttle Thread Ltd and Jamie has also participated in the development of the SmileyCoin wallets.

Countless students have contributed to the tutor-web and the SmileyCoin wallet.

The African Maths Initiative is the Smiley Charity's main partner and they have led the uptake of tablet-based educational technology in Kenya. The work in the western part of Kenya is led by several individuals, particularly Zachariah Mbasu, Thomas Mawora and Maxwell Fundi. The Smiley Charity has a subsidiary in Nairobi, where the library initiative is led by Kamau Mbugua in cooperation with Prof. Evelyn Njurai of Kisii University. The library partners are the Kibera Community Library (internal project leader Mary Kinyanjui), Mathare Community Library (internal project leader Billian Ojiwa) and Kongoni community library (internal project leader Elphas Ongong'o).

## REFERENCES

[1] G. Stefansson, 2004. The tutor-web: "An educational system for class-room presentation, evaluation and self-study," *Computers&Education*, 43 (4): 315-343.

[2] G. Stefansson and A. H. Jonsdottir 2015. "Design and analysis of experiments linking on-line drilling methods to improvements in knowledge," *J. of Statist. Sci, and Applic.* 3(5-6), 63-73.

[3] G. Stefansson and A. J. Sigurdardottir 2009. "Interactive quizzes for continuous learning and evaluation," *In Joint Statistical Meetings (JSM) proceedings* 4577-4591.

[4] G. Stefansson and J. Lentin 2017. "From smileys to Smileycoins: Using a cryptocurrency for rewards in education," *Ledger* vol. 2, 38-54. Retrieved from http://ledgerjournal.org/ojs/index.php/ledger/article/view/103/84

[5] G. Stefansson, A.H. Jonsdottir, TH. Jonmundsson, G.S. Sigurdsson, I.L. Bergsdottir, 2021. "Identifying rote learning and the supporting effects of hints in drills." *INTED 2021*.

[6] G. Stefansson, A.H. Jonsdottir 2021. Learning and evaluation without access to schools during COVID-19. *INTED 2021*.